\documentclass[twocolumn,prd,groupedaddress,10pt]{revtex4}
\usepackage{graphicx}
\usepackage{dcolumn}
\usepackage{bm}

\def\lsim{\mathrel{\mathstrut\smash{\ooalign{\raise2.5pt\hbox{$<$}\cr\lower2.5pt\hbox{$\sim$}}}}}
\def\gsim{\mathrel{\mathstrut\smash{\ooalign{\raise2.5pt\hbox{$>$}\cr\lower2.5pt\hbox{$\sim$}}}}}

\def\be{\begin{equation}}
\def\ee{\end{equation}}
\def\bea{\begin{eqnarray}}
\def\eea{\end{eqnarray}}

\begin{document}

\title{Super-acceleration as Signature of Dark Sector Interaction}

\author{Subinoy Das$^{1,2}$, Pier Stefano Corasaniti$^1$ and Justin Khoury$^3$}
 
\affiliation{$^1$ISCAP, Columbia University, New York, NY 10027, USA \\
$^2$Center for Cosmology and Particle Physics, Department of Physics, New York University, New York, NY 10003, USA\\
$^3$Center for Theoretical Physics, Massachusetts Institute of Technology, Cambridge, MA 02139, USA} 

\begin{abstract} 

We show that an interaction between dark matter and dark energy generically results in an effective dark energy equation of state of $w<-1$. This arises because the interaction alters the redshift-dependence of 
the matter density. An observer who fits the data treating the dark matter as non-interacting will 
infer an effective dark energy fluid with $w<-1$. We argue that the model is consistent with all 
current observations, the tightest constraint coming from estimates of the matter density at 
different redshifts. Comparing the luminosity and angular-diameter distance relations with $\Lambda$CDM 
and phantom models, we find that the three models are degenerate within current uncertainties 
but likely distinguishable by the next generation of dark energy experiments.

\end{abstract}

\maketitle 
 
\section{Introduction} 
\label{intro} 

Nature would be cruel if dark energy were a cosmological constant. Unfortunately this daunting possibility is 
increasingly likely as observations converge towards an equation of state of $w=-1$. 
Combining galaxy, cosmic microwave background (CMB) and Type Ia supernovae (SNIa) data, Seljak {\it et al.}~\cite{seljak} 
recently found $-1.1 \lsim w \lsim -0.9$ at $1\sigma$. On the one hand, a cosmological constant 
is theoretically simple as it involves only one parameter. However, 
observations would offer no further guidance to explain its minuteness, 
whether due to some physical mechanism or anthropic reasoning~\cite{weinberg}.

A more fertile outcome is $w\neq -1$. This implies dynamics --- the vacuum energy is changing in 
a Hubble time --- and hence, new physics. A well-studied candidate is 
quintessence~\cite{quint,quintcmb}, a scalar field $\phi$ rolling down a 
self-interaction potential $V(\phi)$. Its equation of state,
\be
w_{\phi}=\frac{\dot{\phi}^{2}/2-V(\phi)}{\dot{\phi}^{2}/2+V(\phi)}\,,
\label{wphi0}
\ee
can be $<-1/3$ for sufficiently flat $V(\phi)$ and thus lead to cosmic speed-up.
Whether dark energy is quintessence or something else, this case offers hope that further observations, 
either cosmological or in the solar system, may unveil the underlying microphysics of the new sector.

An even more exciting possibility is $w<-1$. In fact there are already indications of this~\cite{alam,coras1} from various independent analyses of the ``Gold'' SNIa dataset~\cite{Riess2}.
Moreover, by constraining redshift parameterizations of $w(z)$ they also exclude that this could 
result from assuming a constant $w$ \cite{Maor,csaki}. The $w<-1$ regime would rule out quintessence since 
$w_{\phi}\ge -1$ (see Eq.~(\ref{wphi0})), as well as most dark energy models.

Devising consistent models with $w<-1$ has proven to be challenging. Existing theories generally involves 
ghosts, such as phantom models~\cite{kamion}, resulting in instabilities and other 
pathologies~\cite{carroll03}. Fields with non-minimal couplings to gravity, 
such as Brans-Dicke theory, can mimic $w<-1$~\cite{BD}. However, solar-system constraints render
the Brans-Dicke scalar field nearly inert, thereby driving 
$w$ indistinguishably close to $-1$. Other proposals for $w<-1$ include brane-world scenarios~\cite{varun}, quantum effects~\cite{onemli}, quintessence-moduli interactions~\cite{chung}, and photon-axion conversion~\cite{Nemanja}. 

In this paper we show that $w<-1$ naturally arises if quintessence interacts with 
dark matter. The mechanism is simple. Due to the interaction, the mass of dark matter 
particles depends on $\phi$. Consequently, in the recent past the dark matter energy density 
redshifts more slowly than the usual $a^{-3}$, which, for fixed present matter density, implies a smaller matter density in the past compared to normal cold dark matter (CDM).

An observer unaware of the interaction and fitting the data assuming normal CDM implicitly 
ascribes this dark matter deficit to the dark energy. The effective dark energy 
fluid thus secretly receives two contributions: the quintessence part and the deficit in dark matter. 
The latter is growing in time, therefore causing the effective dark energy density
to also increase with time, hence $w <-1$. 

Treating dark matter as non-interacting is a {\it sine qua non} for inferring $w<-1$.
There are no wrong-sign kinetic terms in our model --- in fact the combined dark matter plus dark energy 
fluid satisfies $w >-1$. Hence the theory is well-defined and free of instabilities.

Interacting dark matter/dark energy models have been studied in various 
contexts~\cite{dmde,amendola,cham,chamcosmo,wei,peebles04,neil}. Huey and Wandelt~\cite{huey} 
realized that coupled dark matter/quintessence can yield an effective $w<-1$. 
(See also~\cite{stefan} for similar ideas.)
However, the dynamics in~\cite{huey} are such that DM density becomes negligibly small for $z\gsim 1$, thereby forcing the addition of a second non-interacting DM component.
In contrast, our model involves a single (interacting) DM component.

Given the lack of competing consistent models, we advocate that measuring $w< -1$ would hint 
at an interaction in the dark sector. More accurate observations could then search for direct 
evidence of this interaction. For instance, we show that the extra attractive force between dark matter 
particles enhances the growth of perturbations and leads to a few percent excess of power on small scales. 
Other possible signatures are discussed below.

\section{Dark Sector Interaction}\label{model}

Consider a quintessence scalar field $\phi$ which couples to the dark matter via,
{\it e.g.}, a Yukawa-like interaction
\be
f(\phi/M_{\rm Pl})\bar{\psi}\psi\,,
\label{yukawa}
\ee
where is $f$ is an arbitrary function of $\phi$ and $\psi$ is a dark matter Dirac spinor.
In order to avoid constraints from solar-system tests of gravity, 
we do not couple $\phi$ to baryons. See~\cite{cham}, however, for an alternative approach. 

In the presence of this dark-sector interaction, the energy density in the dark matter no 
longer redshifts as $a^{-3}$ but instead scales as
\be
\rho_{\rm DM} \sim \frac{f(\phi/M_{\rm Pl})}{a^3}\,.
\label{rhoDM}
\ee
This can be easily understood since the coupling in Eq.~(\ref{yukawa}) implies
a $\phi$-dependent mass for the dark matter particles scaling as $f(\phi/M_{\rm Pl})$.
Since the number density redshifts as $a^{-3}$ as usual, Eq.~(\ref{rhoDM}) follows.

Thus the Friedmann equation reads
\be
3H^2 M_{\rm Pl}^{2} = \frac{\rho_{\rm DM}^{(0)}}{a^3}\frac{f(\phi/M_{\rm Pl})}{f_0} 
+ \rho_{\phi}\,,\label{friedint}
\ee
where $f_0 = f(\phi_0/M_{\rm Pl})$ with $\phi_{0}$ the field value today, and
\be
\rho_{\phi}=\frac{1}{2}\dot{\phi}^2 + V(\phi)
\ee
is the scalar field energy density. With $a=1$ today, $\rho_{\rm DM}^{(0)}$ is 
identified as the present dark matter density. 

Meanwhile, the scalar field evolution is governed by
\be
\ddot{\phi} + 3H\dot{\phi} = -V_{,\phi} - \frac{\rho_{\rm DM}^{(0)}}{a^3}\frac{f_{,\phi}}{f_0}\,.
\label{phieom2}
\ee
This differs from the usual Klein-Gordon equation for quintessence models by the last term on 
the right-hand side, arising from the interaction with dark matter.

The standard approach to constraining dark energy with experimental data assumes 
that it is a non-interacting perfect fluid, fully described by its equation of state, 
$w_{\rm eff}$. 
Given some $w_{\rm eff}(z)$, the evolution of the dark energy density is then determined by 
the energy conservation equation:
\be
\frac{d\rho_{\rm DE}^{\rm eff}}{dt}=-3H(1+w_{\rm eff})\rho_{\rm DE}^{\rm eff}\,.
\label{continuity}
\ee
Meanwhile, the dark matter is generally assumed to be non-interacting CDM, resulting in the
Friedmann equation
\be
3H^2 M_{\rm Pl}^{2} = \frac{\rho_{\rm DM}^{(0)}}{a^3} + \rho_{\rm DE}^{\rm eff}\,.
\label{fried}
\ee

An observer applying these assumptions to our model would infer an effective dark energy fluid 
with
\be
\rho_{\rm DE}^{\rm eff} \equiv  
\frac{\rho_{\rm DM}^{(0)}}{a^3}\left[\frac{f(\phi/M_{\rm Pl})}{f(\phi_0/M_{\rm Pl})}-1\right] + \rho_\phi\,,\label{rhoeffrel}
\ee
obtained by comparing Eqs.~(\ref{friedint}) and~(\ref{fried}). The end result is to
effectively ascribe part of the dark matter to dark energy. Notice that today
the first term vanishes, hence
the effective dark energy density coincides with $\rho_\phi$. In the
past, however, $\phi\ne\phi_0$, and the two differ. In particular, we will find that
the time-dependence of $\rho_{\rm DE}^{\rm eff}$ can be such that $w_{\rm eff}<-1$.

\begin{figure}[ht]
\centering
\includegraphics[width=80mm]{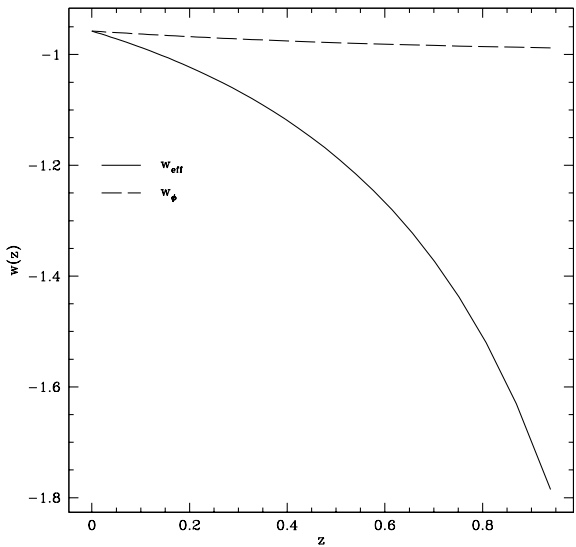}
\caption{Redshift evolution of $w_{\rm eff}$ (solid line) and $w_{\phi}$ (dash line). 
As advocated, $w_{\rm eff}<-1$ in the recent past due to the interaction with the dark matter.}
\label{wefffig}
\end{figure}

To show this explicitly requires an expression for $w_{\rm eff}$. Taking the time derivative of 
Eq.~(\ref{rhoeffrel}) and substituting the scalar equation of motion, Eq.~(\ref{phieom2}), we obtain
\be
\frac{d\rho_{\rm DE}^{\rm eff}}{dt} = 
-3H\left\{\frac{\rho_{\rm DM}^{(0)}}{a^3}\left[\frac{f(\phi/M_{\rm Pl})}{f(\phi_0/M_{\rm Pl})}-1\right] + (1+w_\phi)\rho_\phi\right\}\,.\label{drhoeff2}
\ee
Comparing with Eq.~(\ref{continuity}) allows us to read off $w_{\rm eff}$:
\be
1+w_{\rm eff} = \frac{1}{\rho_{\rm DE}^{\rm eff}}\left\{\left[\frac{f(\phi/M_{\rm Pl})}{f(\phi_0/M_{\rm Pl})}-1\right]\frac{\rho_{\rm DM}^{(0)}}{a^3} + (1+w_\phi)\rho_\phi\right\}.
\label{weff2}
\ee
Now suppose that the dynamics of $\phi$ are such that $f(\phi/M_{\rm Pl})$ increases
in time. This occurs in a wide class of models, as we will see in Section~\ref{phidyn}.
In this case,
\begin{eqnarray}
x\equiv -\frac{\rho_{\rm DM}^{(0)}}{a^3\rho_\phi}\left[\frac{f(\phi/M_{\rm Pl})}{f(\phi_0/M_{\rm Pl})}-1\right]\geq 0
\label{xdef}
\end{eqnarray}
for all times until today, with equality holding at the present time. 
It is straightforward to show that $w_{\rm eff}$ takes a very simple form when expressed in terms of $x$:
\begin{eqnarray}
w_{\rm eff} = \frac{w_\phi}{1-x}\,.
\label{weff3}
\end{eqnarray}
This is our main result. Since $x=0$ today, one has $w_{\rm eff}^{(0)}=w_{\phi}^{(0)}$, 
which is greater than
or equal to $-1$. In the past, however, $x>0$. Moreover, for sufficiently flat potentials, 
$w_\phi\approx -1$. Hence it is possible to have $w_{\rm eff}<-1$ in the past. This is shown 
explicitly in Fig.~\ref{wefffig} for a fiducial case: 
$f(\phi/M_{\rm Pl}) = \exp(\beta\phi/M_{\rm Pl})$ and $V(\phi)=M^4(M_{\rm Pl}/\phi)^\alpha$.

\section{Quintessence dynamics} \label{phidyn}

We now come back to the equation of motion for $\phi$, Eq.~(\ref{phieom2}), and demonstrate that its dynamics can lead to $w_{\rm eff}<-1$. The scalar potential $V(\phi)$ is assumed to satisfy the tracker condition~\cite{wang},
\be
\Gamma\equiv \frac{V_{,\phi\phi}V}{V_{,\phi}^2}> 1\,.
\label{Gamma}
\ee
For an exponential potential, 
$\Gamma=1$, while $\Gamma = 1+\alpha^{-1}$ for $V(\phi)\sim \phi^{-\alpha}$. 
Moreover, we take the coupling function $f$ to be monotonically increasing.

Without coupling to dark matter, the scalar field would run off to infinite values.
Here, however, the interaction has a stabilizing effect since $\phi$ wants to minimize 
the effective potential
\be
V^{\rm eff} = V(\phi) + \frac{\rho_{\rm DM}^{(0)}}{a^3}\frac{f(\phi/M_{\rm Pl})}{f(\phi_0/M_{\rm Pl})}\,.
\ee
Indeed, it is easily seen that the right-hand side of Eq.~(\ref{phieom2}) is just $-V^{\rm eff}_{,\phi}$. Similar stabilization mechanisms have been explored in other contexts, 
such as so-called VAMPS scenarios~\cite{anderson}, string moduli~\cite{pol,others}, 
chameleon cosmology~\cite{cham,chamcosmo}, 
interacting neutrino/dark energy models~\cite{neil}, 
and other interacting dark matter/dark energy models~\cite{huey,anupam}, to name a few.

Having $\phi$ at the minimum of the effective potential is an attractor solution~\cite{chamcosmo}: 
as the dark matter density redshifts due to cosmic expansion, 
$\phi$ adiabatically shifts to larger field values, always minimizing $V^{\rm eff}$. 
This is because the period of oscillations about the minimum, 
$m^{-1}$, is much shorter than a Hubble time, {\it i.e.}, $m\gg H$. 
We show this for the present epoch, leaving the proof for all times as a straightforward exercise.

The mass of small fluctuations about the minimum is given as usual by
\be
m^2=V^{\rm eff}_{,\phi\phi}=\frac{\rho_{\rm DM}^{(0)}}{a^3}\frac{f_{,\phi\phi}}{f_0} 
\left\{1+\frac{f_{,\phi}^2}{f_{,\phi\phi}f}\frac{\Gamma}{V}\frac{\rho_{\rm DM}^{(0)}}{a^3}\frac{f}{f_0}\right\}\,,
\label{m2}
\ee
where we have substituted $\Gamma$ using its definition, Eq.~(\ref{Gamma}).
Evaluating this today, and noting that $\rho_{\rm DM}^{(0)}=3H_0^2M_{\rm Pl}^2\Omega_{\rm DM}^{(0)}$ 
and $V(\phi_0) < 3H_0^2M_{\rm Pl}^2\Omega_{\rm DE}^{(0)}$, we find
\be
\frac{m^2_0}{H_0^2} > 3\Omega_{\rm DM}^{(0)}M_{\rm Pl}^2\left(\frac{f_{,\phi\phi}}{f}\right)_0\left\{1+\Gamma\left(\frac{f_{,\phi}^2}{f_{,\phi\phi}f}\right)_0\frac{\Omega_{\rm DE}^{(0)}}{\Omega_{\rm DM}^{(0)}}\right\}.
\ee
The right hand side is greater than unity for $M_{\rm Pl}^2f_{,\phi\phi}/f\gsim 1$.
In addition, as we will see later, $\Gamma \gg 1$ for consistency with observations of large-scale structure.
These conditions guarantee that fluctuations about the minimum of the effective potential are 
small at the present time. For concreteness, 
let us evaluate this in the case of $f(\phi) = \exp(\beta\phi/M_{\rm Pl})$ 
and $V(\phi)=M^4(M_{\rm Pl}/\phi)^\alpha$:
\be
\frac{m^2_0}{H_0^2} > 3\beta^2\Omega_{\rm DM}^{(0)}
\left(1+\frac{\alpha+1}{\alpha}\frac{\Omega_{\rm DM}^{(0)}}{\Omega_{\rm DE}^{(0)}}\right)\,.
\label{m20}
\ee
This is indeed larger than unity for $\alpha\lsim 1$ and $\beta\gsim {\cal O}(1)$, 
the latter corresponding to a gravitational-strength interaction between dark matter and dark energy.

Next we show that the field is slow-rolling along this attractor solution.
The proof is again straightforward. Differentiating the condition at the minimum, 
$V^{\rm eff}_{,\phi}=0$, with respect to time, we obtain
\be
\dot{\phi} = \frac{3H}{m^2}\frac{\rho_{\rm DM}^{(0)}}{a^3}\frac{f_{,\phi}}{f_0} = -\frac{3H}{m^2}V_{,\phi}\,,\label{pd}
\ee
where in the last step we have used $V^{\rm eff}_{,\phi}=0$.
Thus, 
\be
\frac{\dot{\phi}^2}{2V} = 
\frac{9H^2}{2m^4}\frac{V_{,\phi}^2}{V} <  \frac{9H^2}{2m^2}\frac{1}{\Gamma}\,.
\label{sl}
\ee
Since $m > H$ along the attractor, and since $\Gamma\gg 1$ as mentioned earlier, Eq.~(\ref{sl})
implies that $\phi$ has negligible kinetic energy compared to potential energy, which is the
definition of slow-roll.

The slow-roll property has many virtues. First of all, it implies that 
our attractor solution is different than that derived by Amendola and 
collaborators~\cite{amendola}. In their case, during the matter-dominated era, 
the scalar field kinetic energy dominates over the potential energy and remains a fixed fraction 
of the critical density. This significantly alters the growth rate of perturbations. 
Microwave background anisotropy then constrains the dark matter-dark energy coupling 
to be less than gravitational strength: $\beta < 0.1$ for $f(\phi) = \exp(\beta\phi/M_{\rm Pl})$. 
In our case, as we will see in Sec.~\ref{pert}, 
slow-roll implies a nearly identical growth rate to that in CDM models, even in the 
interesting regime $\beta\gsim 1$.

More importantly, slow-roll means $w_\phi\approx -1$. As argued below Eq.~(\ref{weff3}),
this facilitates obtaining $w_{\rm eff}<-1$.

In essence, slow-roll is enhanced by the dark matter interaction term in Eq.~(\ref{phieom2}) 
which acts to slow down the field. To see this explicitly, note that in usual quintessence models 
(without dark matter interaction), slow-roll is achieved in the large $\Gamma$ limit, for which
\be
\frac{\dot{\phi}^2}{2V} \approx \frac{1}{4\Gamma}\,.
\ee
Comparison with Eq.~(\ref{sl}) shows that this ratio is further 
suppressed by $H^2/m^2\ll 1$ in our case.

The attractor solution described here has a large basin of attraction. The covariant form of Eq.~(\ref{phieom2}) involves $T^\mu_\mu$, the trace of the stress tensor of all fields coupled to $\phi$. These do not exclusively consist of DM. For instance, in a supersymmetric model where the DM is the lightest supersymmetric particle, $\phi$ could conceivably couple to a host of superpartners. Deep in the radiation-dominated era, the $T^\mu_\mu$ source term is generally negligible compared to the Hubble damping term, $3H\dot{\phi}$. However, they become comparable for about a Hubble time whenever a particle species coupled to $\phi$ becomes non-relativistic~\cite{pol}, therefore driving $\phi$ towards the minimum of its effective potential. This provides an efficient mechanism for reaching the attractor~\cite{chamcosmo}.

\section{An explicit example}\label{example}

In this Section we illustrate our mechanism within a specific model.
We consider an inverse power-law potential, $V(\phi)=M^4(M_{\rm Pl}/\phi)^\alpha$, where 
the mass scale $M$ is tuned to $\sim 10^{-3}$~eV in order for acceleration to occur 
at the present epoch. This potential is a prototypical example of a tracker potential
in quintessence scenarios. Its runaway form is in harmony with 
non-perturbative potentials for moduli in supergravity and string theories. 

The coupling function is chosen to be $f(\phi) = \exp(\beta\phi/M_{\rm Pl})$. 
The exponential form is generic in dimensional reduction in string theory 
where $\phi$ measures the volume of extra dimensions. 
Moreover, $\beta$ is expected to be of order unity, corresponding to gravitational strength.
While the coupling to matter exacerbates the fine-tuning of the quintessence potential~\cite{carroll98},
we find the phenomenological consequences of our model sufficiently interesting to
warrant sweeping naturalness issues under the rug. 

In this example, the condition at the minimum reads
\be
-\frac{\alpha M^4M_{\rm Pl}^\alpha}{\phi^{\alpha+1}} 
+ \frac{\beta}{M_{\rm Pl}} \frac{\rho_{\rm DM}^{(0)}}{a^3} e^{\beta (\phi-\phi_0)/M_{\rm Pl}} =0\,.
\label{minfid}
\ee
Evaluating this today, and noting that $V_0\approx 3H^2_0M_{\rm Pl}^2\Omega_{\rm DE}^{(0)}$
because of slow-roll, we obtain
\be
\frac{\phi_0}{M_{\rm Pl}} \approx \frac{\alpha}{\beta} \frac{\Omega_{\rm DE}^{(0)}}{\Omega_{\rm DM}^{(0)}}\,.
\label{phi0}
\ee
Equations~(\ref{minfid}) and~(\ref{phi0}) combine to provide a simple expression for the 
redshift-evolution of 
$\phi$ as it follows the minimum of the effective potential:
\be
\left(\frac{\phi}{\phi_0}\right)^{\alpha+1} = (1+z)^{-3}e^{\beta(\phi_0-\phi)/M_{\rm Pl}}\,.
\label{phioverphi0}
\ee

Next we calculate the resulting effective equation of state. To do so, we first need an 
expression for $\rho_\phi$ as a function of redshift. Notice that in the slow-roll approximation, $\rho_{\phi}\approx V(\phi)$.
This does not imply, however, that $\rho_{\phi}\approx {\rm const.}$, since $\rho_{\phi}$ 
does not obey the usual conservation equation. Using Eq.~(\ref{minfid}), we instead have
\be
\rho_\phi\approx \frac{V}{V_{,\phi}}V_{,\phi}= 
\frac{\beta}{\alpha}\frac{\phi}{M_{\rm Pl}}\frac{\rho_{\rm DM}^{(0)}}{a^3}e^{\beta (\phi-\phi_0)/M_{\rm Pl}}\,.
\label{r}
\ee
Substituting this and Eq.~(\ref{r}) in the definition of $x$ given in Eq.~(\ref{xdef}), we arrive at
\be
x= \frac{\Omega_{\rm DM}^{(0)}}{\Omega_{\rm DE}^{(0)}}\frac{\phi_0}{\phi}\left\{\exp\left[\alpha\frac{\Omega_{\rm DE}^{(0)}}{\Omega_{\rm DM}^{(0)}}\left(1-\frac{\phi}{\phi_0}\right)\right]-1\right\}\,.
\label{x1}
\ee
This shows explicitly that $x$ is a positive, monotonically increasing 
function of $z$ which vanishes today. Moreover, since the field is slow-rolling, 
we have $w_\phi\approx -1$. Therefore, Eq.~(\ref{weff3}) implies
\be
w_{\rm eff} \approx -\frac{1}{1-x} \leq -1\,,
\ee
with the approximate equality holding today. Hence this yields an effective dark energy 
fluid with $w<-1$ in the recent past.

Note from Eq.~(\ref{x1}) that $x=1$ at some time in the past, implying 
that $|w_{\rm eff}|$ momentarily diverges and then becomes positive again at higher redshifts. 
This is because $\rho^{\rm eff}_{\rm DE}$ eventually becomes negative, at which point 
the effective dark 
energy fluid has both negative pressure and energy density. 
As $z$ increases further and $x$ becomes large, one has $w_{\rm eff}\approx 0$, and the 
fluid behaves like dust.

In Fig.~\ref{wefffig} we plot the redshift evolution of $w_{\rm eff}$ and $w_{\phi}$ 
for $\alpha=0.2$, $\beta=1$ and $\Omega_{\rm DE}^{(0)}=0.7$. 
(As will be discussed in the next Section, a small value for $\alpha$ is required for 
consistency with large-scale structure observations.) While $w_{\phi}$ remains bounded 
from below by -1, $w_{\rm eff}$ is less than -1 for $z\gsim 0.1$, as claimed above. 

The evolution of $w_{\rm eff}(z)$ shown in Fig.~\ref{wefffig} is consistent with the observational limits
on redshift dependent parameterizations of the dark energy equation of state~\cite{coras1}.
One way to see this is to consider the weighted average 
\be
\bar{w}_{\rm eff}\equiv\frac{\int \Omega_{\rm eff}(a)w_{\rm eff}(a)da}{\int \Omega_{\rm eff}(a)da}\,,
\ee
where the integral runs from $z=0$ up to the maximum redshift of current SN Ia data, $z\sim1.5$.
This gives $\bar{w}_{\rm eff}\approx -1.1$, which lies within the allowed range of $w$ found in~\cite{seljak}.
Note that while Fig.~\ref{wefffig} was derived using the above analytical expressions, we have
checked these against numerical solutions of the equations of motion and found very good agreement.

\section {Observational Constraints and Consequences}\label{obs}

We have shown that the interaction between quintessence and dark matter can mimic the cosmology of a phantom fluid. 
In this Section we discuss some observational consequences of this scenario and 
argue that it is consistent with current observations. 
At the level of homogeneous cosmology this is certainly true, as long as parameters are chosen
such that $w_{\rm eff}$ lies within the allowed range.
We argue that this is also the case when considering inhomogeneities, at least at the linear
level. The main effect here is the fifth force between dark matter particles mediated by $\phi$, which
enhances the growth rate of density perturbations.
 
A rigorous comparison with observations would require a full likelihood analysis 
including a host of cosmological probes, which is beyond the scope of this paper. 
We instead contend ourselves with a simplified (and perhaps more conservative) analysis 
to derive general constraints. As in Sec.~\ref{example}, we focus on an exponential 
coupling function and inverse power-law potential.

\subsection{Mass estimates from large-scale structure}

The tightest constraint comes from various estimates of the dark matter density at different redshifts.
Since the dark matter redshifts more slowly than $a^{-3}$ in our model, then for fixed present 
matter density this implies a smaller matter density in the past compared to a CDM model. 
Indeed, at early times ($\phi\ll \phi_0$), the matter density differs from that of a usual 
dust CDM model by
\be
\frac{\rho_{\rm DM}}{\rho_{\rm CDM}} \rightarrow e^{-\beta\phi_0/M_{\rm Pl}} = \exp\left(-\alpha\frac{\Omega_{\rm DE}^{(0)}}{\Omega_{\rm DM}^{(0)}}\right)\,,
\label{ratrho2}
\ee
where in the last step we have used Eq.~(\ref{phi0}). 

It is reasonable to assume that this ratio cannot deviate too much from unity, for 
otherwise we risk running into conflict with 
estimates of the matter density at various redshifts, {\it e.g.} from galaxy counts, 
Lyman-$\alpha$ forest, weak lensing etc. This is supported by the fact that the allowed 
range of $\Omega_{\rm DM}^{(0)}$ is almost independent of the specifics of the dark energy,
as derived from a general analysis~\cite{kunz,coras1} of the combined SNIa ``Gold'' data~\cite{Riess2}, Wilkinson Anisotropy Microwave Probe (WMAP) power spectra~\cite{peiris} and Two-Degree Field 
(2dF) galaxy survey~\cite{2df}.
In particular $0.23\lsim \Delta\Omega_{\rm DM}^{(0)}\lsim 0.33$ at $2\sigma$ (see also~\cite{mortsell,seljak}).
Substituting $\Omega_{\rm DM}^{(0)}=0.33$ in Eq.~(\ref{ratrho2}), we obtain
\be
\alpha\lsim 0.2\,.
\label{condalp}
\ee
Thus dark matter density estimates require the scalar field potential 
to be sufficiently flat, thereby making the attractor behavior and slow-roll condition 
discussed in Sec.~\ref{phidyn} more easily satisfied.

Equation~(\ref{ratrho2}) shows that $\rho_{\rm DM}$ redshifts like normal CDM ({\it i.e.}, $\rho_{\rm DM}\sim a^{-3}$) for most of the cosmological history, except in the recent past. This is crucial in satisfying constraints on $\Omega_{\rm DM}^{(0)}$ and traces back to our choice of inverse power-law potential. In contrast, the exponential potential studied in~\cite{huey} has a very different attractor solution. In this case, dark energy remains a constant fraction of the total energy density and modifies the DM equation of state at all redshift. This in turn renders the matter density negligibly small for $z\gsim 1$. Therefore, in order to satisfy constraints on $\Omega_{\rm DM}^{(0)}$ (as well as $z_{\rm eq}$), one must introduce a second DM component, which is non-interacting and dominates for most of the history.

Finally, we note that while Eq.~(\ref{condalp}) is an extra tuning on $V(\phi)$, normal 
quintessence also suffers from the same constraint. Indeed, ``tracker'' quintessence 
with $V(\phi)=M^4(M_{\rm Pl}/\phi)^\alpha$ leads to a dark energy equation of state
\be
w_{\phi} = -\frac{2}{\alpha+2}\,.
\ee
Imposing the current observational constraint $w<-0.9$ results in a bound 
on $\alpha$ identical to Eq.~(\ref{condalp}).

\subsection{CMB and SNIa observables} \label{cmbsna}

We now focus on cosmological distance tests, in particular the SNIa luminosity-distance relation
and the angular-diameter distance to the last scattering surface as inferred from 
the position of CMB acoustic peaks. We wish to compare these observables 
for three different models, namely the interacting scalar field dark matter model with 
$\alpha=0.2$ and $\beta=1$, a $\Lambda$CDM model, and a phantom model with $w=-1.2$.

\begin{figure}[ht]
\centering
\includegraphics[width=80mm]{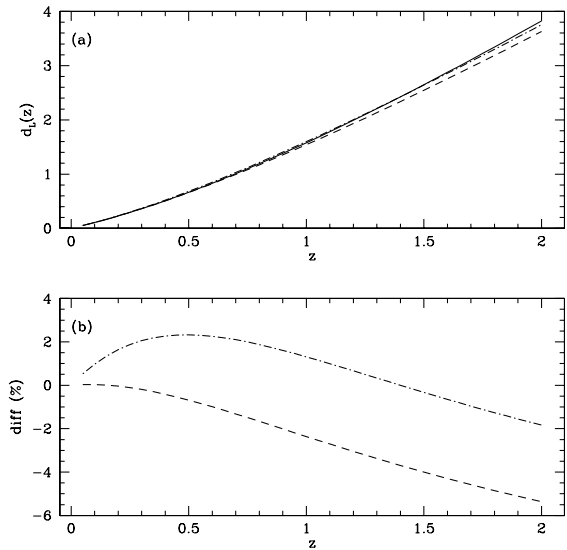}
\caption{Upper panel shows the luminosity distance ($d_L$) as function of redshift 
for our model (solid) a phantom model with $w=-1.2$ (dash-dot) and $\Lambda$CDM (dash). 
We have fixed $\Omega_{\rm DM}^{(0)}=0.3$. Lower panel shows the percentage difference 
between our model and phantom (dash-dot), and between our model and $\Lambda$CDM (dash), respectively.}
\label{dL1}
\end{figure}

The position of Doppler peaks depends on the angular-diameter distance to the 
last scattering surface,
\be
d_{\rm A}(z_{\rm rec}) = (1+z_{\rm rec})^{-1}\int_0^{z_{\rm rec}}\frac{dz}{H(z)}\,,
\ee
where $z_{\rm rec}$ is the redshift at recombination. Observations of SNIa, on the other hand, probe 
the luminosity distance 
\be
d_{\rm L}(z) = (1+z)\int_0^z\frac{dz}{H(z)}\,.
\ee
Figure~\ref{dL1}a shows the luminosity distance for all three models 
with $\Omega_{\rm DM}^{(0)}=0.3$, while Fig.~\ref{dL1}b gives their percentage difference. 
The difference between our model and $\Lambda$CDM is $\lesssim 4\%$ for $z<1.5$;
similarly the difference with respect to the phantom model is within $\lesssim 2\%$. 
Thus all three models are degenerate within the uncertainties of present SNIa data which 
determine $d_L(z)$ to no better than $\sim 7\%$. Furthermore, this suggests that percent-level 
accuracy from future SNIa experiments such as the Supernova Acceleration Probe (SNAP)~\cite{snap}, combined with other cosmological probes, could distinguish between them.

Since $\Omega_{\rm DM}^{(0)}$ is kept fixed in this case, the matter density 
in the interacting dark energy model differs in the past from that in the $\Lambda$CDM 
and phantom cases, as seen from Eq.~(\ref{ratrho2}).
This results in a $10\%$ difference in $d_{\rm A}(z_{\rm rec})$, which is again within 
current CMB uncertainties.

\begin{figure}[ht]
\centering
\includegraphics[width=80mm]{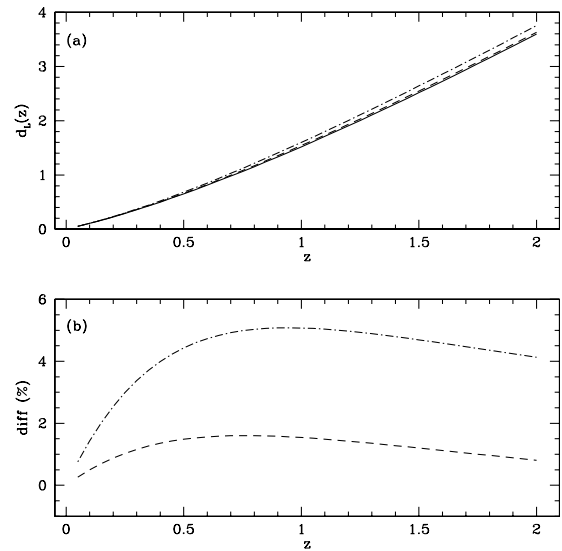}
\caption{Same as in Figure~\ref{dL1}, except $\Omega_{\rm DM}^{(0)}=0.4$ for the interacting scalar 
field dark matter model in this case. This gives equal $d_{\rm A}(z_{\rm rec})$ for all three models.}
\label{dL2}
\end{figure}

Suppose we instead keep $d_{\rm A}(z_{\rm rec})$ fixed, which essentially amounts 
to fixing the matter density at high redshift. With $\Omega_{\rm DM}^{(0)}=0.3$ for both
the $\Lambda$CDM and phantom models, this is achieved by setting $\Omega_{\rm DM}^{(0)}=0.4$ for our 
model. These values are compatible with current limits, as mentioned earlier. 
The resulting luminosity distances and percentage differences are plotted in Fig.~\ref{dL2}. 
In this case we find that our model is more closely degenerate with $\Lambda$CDM than phantom.

\subsection{Growth of density perturbations} \label{pert}

In the slow-roll approximation the evolution equation for dark matter inhomogeneities, $\delta=\delta\rho_{\rm DM}/\rho_{\rm DM}$, is given in synchronous gauge by~\cite{chamcosmo}
\be
\delta'' + aH \delta' = \frac{3}{2}a^2H^2
\left[ 1 + \frac{2\beta^2}{1+a^2 V_{,\phi\phi}/k^2}\right]\delta\,,
\label{densitypertur}
\ee
where primes denote differentiation with respect to conformal time. 
This differs from the corresponding expression in CDM models only through the factor in square brackets, 
normally equal to unity. Since this term accounts for the self-attractive 
force on the perturbation, the extra contribution proportional to $\beta^2$ arises 
from the attractive fifth force mediated by the scalar field.
This force has a finite range, which for an inverse power-law potential is
\be
\lambda = V^{-1/2}_{,\phi\phi} = \sqrt{\frac{\phi^{\alpha+2}}{\alpha(\alpha+1)M^4M_{\rm Pl}^\alpha}}\,.
\label{lambdaz}
\ee

Perturbations with physical wavelength much larger than $\lambda$, {\it i.e.} $a/k\gg \lambda$, evolve as normal CDM. 
On the other hand, perturbations with $a/k\ll \lambda$, evolve as if Newton's constant were 
a factor of $1+2\beta^2$ larger. Thus the interaction with the quintessence field leads to an enhancement of power on small scales~\cite{frieman}. In particular, small-scale perturbations go non-linear at higher redshift than in $\Lambda$CDM, as shown recently in a closely related context of chameleon cosmology~\cite{brax}.
(Numerical simulations have also found that a similar attractive scalar interaction for dark matter particles, albeit with a much smaller range of 1~Mpc, results in emptier voids between concentrations of large galaxies~\cite{peebles}.)

Quantitatively, from Eqs.~(\ref{phi0}) and~(\ref{phioverphi0}) in the limit $\alpha\ll 1$,
we obtain
\be
V_{,\phi\phi} \approx H_0^2(1+z)^6e^{2\beta(\phi-\phi_0)/M_{\rm Pl}}\frac{3\beta^2}{\alpha}\frac{(\Omega_{\rm DM}^{(0)})^2}{\Omega_{\rm DE}^{(0)}}\,,
\ee
where $H_0$ is the present value of the Hubble parameter. This implies, for instance, 
that at the present epoch
\be
\lambda^{(0)} = H_0^{-1}\sqrt{\frac{\alpha\Omega_{\rm DE}^{(0)}}{3\beta^2(\Omega_{\rm DM}^{(0)})^2}}\approx 0.7H_0^{-1}\,,
\label{lambda0}
\ee
where in the last step we have taken $\alpha=0.2$, $\beta=1$ and $\Omega_{\rm DM}^{(0)}=0.3$. Hence the present range of this fifth force is 
comparable to the size of the observable universe. 
However, $\lambda$ varies with redshift, and it is easily seen that $\lambda\ll H^{-1}$ in the past.
In particular, we do not expect measurable 
effects in the CMB. This is in contrast with quintessence models~\cite{quintcmb}, 
as well as the interacting dark matter/dark energy model of Amendola and collaborators~\cite{amendola}, where $m\sim H$ along the attractor solution, leading to imprints in the CMB. 

We solve numerically Eq.~(34) and compute the linear matter power
spectrum, $\Delta^2(k)\propto k^3 P(k)$, normalized to
WMAP~\cite{peiris}, where $P(k)=|\delta_k|^2$. In Fig.~4a we plot
the resulting power spectrum for our model (solid line) and
$\Lambda$CDM (dash line) with $\Omega_{\rm DM}^{(0)}=0.4$ and $0.3$
respectively. The two curves are essentially indistinguishable by eye.

In Fig.~4b we plot the fractional difference between the two spectra.
The discrepancy is $<2\%$ on the scales probed by current
large scale structure surveys and consistent with the experimental
accuracy of 2dF Galaxy Redshift Survey~\cite{2df} and Sloan
Digital Sky Survey (SDSS)~\cite{sdss}. On large scales the
perturbations in the two models evolve in a similar way ($k<0.01~{\rm
h Mpc^{-1}}$), while on intermediate scales ($0.01< k<0.4~{\rm h
Mpc}^{-1}$) the $\Lambda$CDM shows a few percent excess of power which
is mostly due to small difference in the expansion rate of the two
models after decoupling. Most importantly, on smaller scales ($k>0.4~{\rm h Mpc^{-1}}$) the power spectrum of $\Lambda$CDM is suppressed
compared to our model. This is due to the fifth force which
enhances the clustering of dark matter
perturbations compared to the uncoupled case. However, in this range
perturbations become non-linear; hence a rigorous study of how this fifth
force affects structure formation requires N-body simulations.

\begin{figure}[ht]
\centering
\includegraphics[width=80mm]{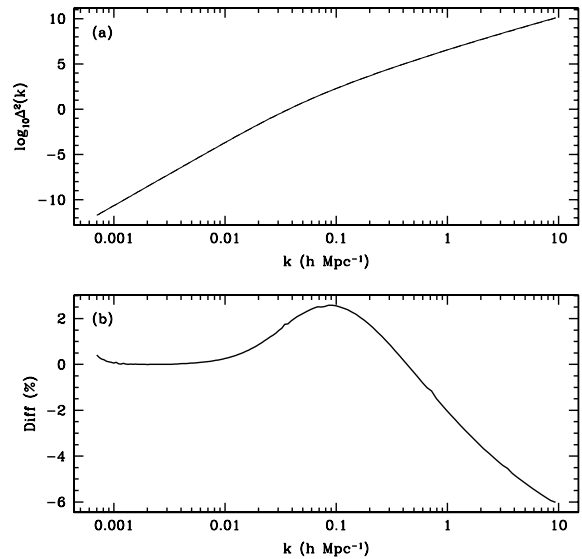}
\caption{Upper panel shows the matter power spectrum ($\Delta^2(k)$) over the relevant range of scales for our model (solid) and $\Lambda$CDM (dash) with $\Omega_{\rm DM}^{(0)}=0.4$ and $0.3$, respectively. Lower panel shows the percentage difference between the two curves, which is well within current experimental accuracy.}
\label{Pk}
\end{figure}

\subsection{Galaxy and cluster dynamics}

Since the $\phi$-mediated force is long-range today (see Eq.~(\ref{lambda0})), our model is subject to constraints from galaxy
and cluster dynamics~\cite{frieman}. For instance, a fifth force in the dark sector leads to a discrepancy
in mass estimates of a cluster acting as a strong lens for a high-redshift galaxy. Lensing measurements probe the actual mass since photons are oblivious to the fifth force, while dynamical observations are affected and would overestimate the mass of the cluster.

Other effects studied in~\cite{frieman} include mass-to-light ratios in the Local Group, rotation curves of galaxies in clusters, and dynamics of rich clusters. These combine to yield a constraint of $\beta\lsim 0.8$, consistent with our assumption of $\beta\sim {\cal O}(1)$. This is consistent with generic string compactifications; if for instance $\phi$ is the radion field measuring the distance between two end-of-the-world branes, $\beta=1/\sqrt{6}$~\cite{chamcosmo}.

\section{Discussion}\label{conclusions}

In this paper we have shown that an interaction between dark matter and dark energy generically mimics $w<-1$ cosmology, provided that the observer treats the dark matter as non-interacting. Unlike phantom models, the theory is well-defined and free of ghosts.

Our model is consistent with current observations provided the scalar potential is sufficiently flat. 
For our fiducial $V(\phi)=M^4/\phi^\alpha$, this translates into $\alpha\lsim 0.2$. 
This is no worse than normal quintessence with tracker potential, where a nearly identical bound follows from observational constraints on $w_\phi$.

In fact our scenario is less constrained than other interacting dark energy/dark matter models studied
in the literature. There is no need to introduce a non-interacting DM component, as in~\cite{huey}; nor does the coupling strength need be much weaker than gravity, $\beta\lsim 0.1$, as in~\cite{amendola}. 
Instead, our model allows for a single interacting DM species with gravitational strength coupling to dark energy --- $\beta\sim {\cal O}(1)$. In both cases this traces back to a difference in attractor solutions.

At the level of current uncertainties, the model is degenerate with both $\Lambda$CDM 
and phantom models. However, our calculations of luminosity and angular-diameter 
distances indicate that these models could be distinguished by the next generation 
of cosmological experiments devoted to the study of dark energy, such as SNAP, the Large Synoptic Survey Telescope (LSST)~\cite{lsst}, the Joint Efficient Dark-energy Investigation (JEDI)~\cite{jedi}, the Advanced Liquid-mirror Probe for Astrophysics, Cosmology and Asteroids (ALPACA)~\cite{alpaca}, and others.

A dark sector interaction may reveal itself in various ways in the data.
A strong hint would be a preference for $w<-1$ when fitting cosmological
distance measurements assuming CDM.
Another indication is a discrepancy between the clustering matter density at various redshifts and the
expected $(1+z)^3$ dependence in normal CDM models, which could appear as a discrepancy in the inferred value of $\Omega_M^{(0)}$.
 
We also uncovered modifications in the linear matter power spectrum and large-scale structure.
These are primarily due to the attractive scalar-mediated force which enhances the growth of DM perturbations on small scales. Note that the opposite behavior obtains for a phantom scalar coupled to dark matter, resulting in a repulsive scalar force which damps perturbations~\cite{amendolaphantom}. As mentioned earlier, non-linear effects
are important for the relevant range of scales and would require N-body simulations.

Other observational effects that could distinguish our model from
$\Lambda$CDM and phantom include the bias parameter. Since baryons are
unaffected by the fifth force, baryon fluctuations develop a
constant large-scale bias~\cite{LucaDome} which could be
observable. Similarly, comparison of the redshift dependence of the
matter power spectrum, $P(k,z)$, may be useful to constrain the scale
$\lambda$, which varies with $z$.
Moreover, it has recently been argued that dark sector interactions could lead to oscillations in 
the redshift dependence of cluster counts~\cite{mota}.
The integrated Sachs-Wolfe effect is another mechanism worth studying. 
Since the present range of our scalar force is comparable to the size of the
observable universe, it might account for the observed lack of
power on large scales in the CMB.

\acknowledgments
We are grateful to L.~Amendola, R.~Caldwell, E.~Copeland, G.~Huey, M.~Trodden, B.~Wandelt and N.~Weiner for insightful comments. S.D. is thankful to L.~Hui for support under DOE grant number DE-FG02-92-ER40699 and
B.~Greene for the opportunity to work at ISCAP. This work is supported in part by the
Columbia Academic Quality Funds (P.S.C.) and the U.S. Department of Energy 
under cooperative research agreement DE-FC02-94ER40818 (J.K.).

\end{document}